\newcommand{\beq}{  \begin{eqnarray}}
\newcommand{\eeq}{  \end{eqnarray}}
\documentclass[showpacs,preprintnumbers,amsmath,amssymb]{revtex4}

\usepackage{graphicx}
\usepackage{dcolumn}
\usepackage{bm}

\begin{document}
\title{Dipolar Relaxation of Cold Sodium Atoms in \\
a Magnetic Field}
\author{B. Zygelman }
\email{bernard@physics.unlv.edu}
\affiliation{
Department
of Physics, University of Nevada Las Vegas, Las Vegas NV 89154, USA}
\affiliation{ MIT-Harvard Center for Ultra-Cold Atoms,
Cambridge MA 02139 USA}
\altaffiliation{Visiting Scientist, 2001}

\begin{abstract}
A quantum mechanical close coupling theory of
spin relaxation in the stretched 
$F=2, M_{F}=2$, hyperfine level of sodium is presented.
We calculate the dipolar relaxation rate of magnetically 
trapped cold sodium atoms in the magnetic field range
$ \rm 0< B < 4 \, Tesla $.
 The influence of shape resonances and the anisotropy of
the dipolar interaction on the collision dynamics
are explored. We examine the sensitivity of the
calculated cross sections on the choice of
asymptotic atomic state basis. We calculate and compare 
elastic scattering with dipolar relaxation cross sections
for temperatures ranging from the ultra-cold to $\rm 2 \,K$.
We find that the value for the ratio of elastic to inelastic
cross sections favor application of proposed buffer gas cooling
and loading schemes.
\end{abstract}

\maketitle

\section{introduction}
Advances in the  cooling and trapping of atoms have 
greatly facilitated the exploration of quantum degenerate matter \cite{leg01}.
 The realization of
 Bose-Einstein condensation (BEC) \cite{and95,dav95,brad95,fried98}
 in atomic vapors validates the
standard theory \cite{huang57}
 for non-interacting and weakly interacting systems, but
 experiments \cite{kett98,cour98,rob98}
 demonstrate that atomic interactions,
 though weak in an ensemble of atoms in the gas phase, lead to 
 interesting phenomena that are not present in the ideal gas system
  \cite{moer95,kett99,rob00,don01}.

In a dilute, cold, gas, atoms interact primarily via long range
dispersion and exchange forces. However, inelastic processes
are driven by spin exchange \cite{dalg61,zyg03}
 and dipolar interactions \cite{ver86}. The latter 
process does not conserve total, atom pair, spin
angular
momentum
 and it is a primary mechanism by which
atoms, having hyperfine structure, can suffer an
inelastic transition. Dipolar relaxation determines the lifetime
of the hydrogen atom BEC \cite{grey00}, 
contributes to heating
and influences the operation of atomic clocks
\cite{kok97} .
Rates for dipolar relaxation have been measured in
 $ \rm ^{7}Li$ \cite{hul99}, $\rm Cs $ \cite{sod98}, $\rm ^{85} Rb$
 \cite{rob00}, 
$\rm H $ \cite{grey00} and $\rm Cr$ \cite{wein01}.
 The rates are generally small,
 typically having values that range $ 10^{-14} - 10^{-16} \, cm^{3} s^{-1}$, 
 but in the cases of $\rm Cs $ and $\rm Cr$ 
 anomalously large values  
 have been reported \cite{sod98,wein01,decarv03}. Calculations
 for dipolar relaxation rates, in the zero temperature limit,
 of several species have also been 
 reported \cite{stoof88,ver86,ver91,ver93,ver96,mies96,mies00}. 

 New magnetic trapping and buffer gas cooling schemes \cite{doy99,harris03}
 create the
 opportunity to study a host of atomic and molecular
 species that are not amenable to laser cooling technology. 
 In a typical loading scheme, species with large magnetic moments are
 trapped by external fields at relatively high temperatures,
 on the order of 1 K, before they are cooled into the sub-Kelvin
 regime. In order to model this process one needs a detailed
 understanding 
 of the collision processes that can lead to trap loss
 and heating. To that end, we present a comprehensive quantum
 mechanical theory of dipolar relaxation in alkali atoms. The
 theory is suited for application in
 gases at temperatures where many partial
 waves in the collision wave function contribute, and is applicable
 at arbitrary external magnetic field intensity. We apply the theory to
 calculate the dipolar relaxation rate of the  stretched hyperfine
 level in the $\rm  ^{23} Na(3s)$ atom and, in this paper, present results for
 temperatures that range from the ultra-cold to several Kelvin
 and magnetic field intensities in the range $ \rm 0 < B < 4 \,\rm Tesla $. The results of our calculation
 are compared to previous theoretical predictions\cite{ver91,ver96}.
 We present, the first fully quantum mechanical calculation for dipolar
 relaxation of sodium atoms in a magnetic field at higher temperatures where
 many partial waves contribute.
 We  give a detailed
 description of the collision theory and explore the consequences
 of anisotropy on the collision
 dynamics. We point out the importance of shape resonances 
 and their influence on the value of
 the inelastic rate. We identify a feature in the cross section
 that corresponds to the presence of an above threshold resonance, or
 a virtual state,  in the $ l=2 $ partial wave of 
 the scattering amplitude.

 In section I we provide an introduction to the theoretical
 formalism that is applied in the calculations. A detailed
 discussion of the close coupling equations, asymptotic
 boundary conditions and symmetry requirements is given
 in sections II, and III. In section IV we present the
 results of our calculations, and provide a detailed analysis
 of these results. Unless it is otherwise stated, atomic units
 are used throughout the discussion.
 
\section{Channel basis}

We consider  two sodium atoms  in their
$ F=2,\, M_{F}=2$ hyperfine level, the maximal stretched state.
In Table I we itemize two-atom hyperfine levels with the
notation $ |F M_{F} f_{a} f_{b}> $, where $F$ is the total
angular momentum of the two atoms, $ M_{F}$ the azimuthal
projection of that angular momentum, and $ f_{a} $, $ f_{b} $
are the total angular momenta for atoms  $a$ and $b$ respectively.
The states listed can mix, through dipolar and spin exchange
interactions, with the maximal stretched state
 during
a collision.
\begin{table} 
\caption{Quantum numbers associated with the various basis representations.
$M_{F}$ is the total spin angular momentum along the quantization axis and
 $\Delta_{F} $ is the energy defect between the $ F=2$ and $F=1$ hyperfine
levels in Sodium.}
\begin{ruledtabular}
 \begin{tabular}{c|l|c|c|c|c|c}
  $ M_{F} $ & index &  $ \vert f_{a} m_{a}  f_{b}   m_{b}> $  & level &
   Energy &   $\vert F M_{F} f_{a} f_{b}> $    &  $ \vert S   M_{S} I   M_{I} >
$ \\
\hline
 4 & 1& 2 2 2 2 & h h &   $ 2 \Delta_{F} $  &4 4 2 2   &1 1 3 3  \\
\hline
 3 & 2& 2 2 2  1 & h g &  $ 2 \Delta_{F} $   &4 3 2 2   & 1 1 3 2\\
 3 & 3 &2 1 2  2 & h g &   $ 2 \Delta_{F} $  &3 3 2 2  & 1 0 3 3 \\
 3 & 4 &2 2 1  1 & h a &    $  \Delta_{F} $ &3 3 2 1  & 1 1 2 2\\
 3 & 5 &1 1 2  2 & h a &   $  \Delta_{F} $  &3 3 1 2  & 0 0 3 3 \\
\hline
 2 & 6 &2 2 2 0 & h f  &   $ 2 \Delta_{F} $  &4 2 2 2   & 1 1 3 1    \\
 2 & 7 &2 0 2 2  & h f &   $ 2 \Delta_{F} $  &3 2 2 2   & 1 0 3 2  \\
 2 & 8 &2 2 1 0  & h b &    $  \Delta_{F} $  &3 2 2 1   & 1 1 2 1  \\
 2 & 9 &1 0 2 2  & h b &    $  \Delta_{F} $  &3 2 1 2   & 1 0 2 2 \\
 2 & 10 &2 1 2  1 & g g &   $ 2 \Delta_{F} $  &2 2 2 2  & 1 -1 3 3 \\
 2 & 11 &1 1 2  1 & g a &    $  \Delta_{F} $ &2 2 1 2  & 1 1 1 1 \\
 2 & 12 &2 1 1  1 & g a &    $  \Delta_{F} $ &2 2 2 1  & 0 0 3 2 \\
 2 & 13 &1 1 1  1 & a a &    $  0 $          &2 2 1 1  & 0 0 2 2 \\
\end{tabular}
\end{ruledtabular}
\end{table}

 Dipolar interaction selection rules(discussed in the
sections below), allow a change in the azimuthal quantum number
$ \Delta_{M_{F}} =2,1 $ and thus the states itemized in Table
I must be included in the close coupling expansion.
 States within
a given $ M_{F}$ manifold can undergo spin-exchange transitions.
 In Table
I we also list the basis $  |f_{a} m_{a}  f_{b} m_{b}> $
in which the individual atom azimuthal
angular momenta are good quantum numbers. The basis 
$ | S M_{S} I M_{I} > $ diagonalizes the asymptotic Hamiltonian
  if the hyperfine interaction
can be neglected, i.e. at large magnetic field strengths. Here,
$ S, \, M_{S} $ are the total two-atom spin angular momentum
quantum numbers, and $ I, M_{I} $ the nuclear angular momentum quantum
numbers. Allowed values for $ \rm ^{23} Na_{2} $
are $ S=1,0$ and $ I=3,2,1,0 $.
In the close coupling expansion involving molecular channel
states we keep the notation that is
appropriate for the asymptotic region to itemize the states in
the expansion. For example, we define molecular channel basis
 $ |S M_{S} I M_{I}> \equiv
 |^{3} \Sigma_{u}> \otimes |I M_{I}> $ for $ S=1$, and 
$ |S M_{S} I M_{I}> \equiv   |^{1} \Sigma_{g}> \otimes|I M_{I}> $ for $S=0$,
where $ | ^{3} \Sigma_{u}>, |^{1} \Sigma_{g}> $ are Born-Oppenheimer (BO)
eigenstates for the ground $ \rm Na_{2}$ system.
The molecular channel basis merge to the correct
asymptotic basis at large inter-nuclear separation. The states
$ |F M_{F} f_{a} f_{b}> $ are then defined by the linear combination
of the BO channel states given above, 
\beq
  |F M_{F} f_{a} f_{b}> \equiv \sum 
 |S M_{S} I M_{I}><M_{I} I M_{S} S  |F M_{F} f_{a} f_{b}>
\label{qqq1}
\eeq 
where the coefficients $ <M_{I} I M_{S} S  |F M_{F} f_{a} f_{b}> $
are standard recoupling coefficients appropriate for the 
asymptotic basis. In this way we insure  the molecular
close coupling expansion accounts for the asymptotic hyperfine 
interaction within each atom.
In the case of a homonuclear system, the basis $ |F M_{F} f_{a} f_{b}>$
is not an eigenstate of the electron inversion operator, but we
can define the states $ |F M_{F} (f_{a} f_{b}) > \equiv 
{1 \over \sqrt{2}} (|F M_{F} f_{a} f_{b}> \pm |F M_{F} f_{b} f_{a}>$,
that are eigenstates.

In a non-zero magnetic field, the asymptotic Hamiltonian is
not diagonal in the representation defined by the basis vectors
introduced above.
In addition to asymptotic
hyperfine interactions, each atom experiences the Zeeman interaction.
In that case, a linear combination of the basis states defined above must be
found so that the asymptotic Hamiltonian is diagonal in the new
representation. We express these states using the notation
$ |M_{F} \, p \, \epsilon_{i} > $, where $ M_{F}$ is the total angular
momentum along the direction fixed by the magnetic field, $ p$ is
an inversion parity quantum number, and $\epsilon_{i}$ is the
asymptotic energy level eigenvalue.
 We discuss the construction of that basis in the sections
below.

In Figs. 1a,1b we illustrate the energy spectrum of
the asymptotic Hamiltonian for the $ \rm Na_{2}$ system,
as a function of magnetic field strength. 

\begin{figure}
\includegraphics[scale=.5,angle=0]{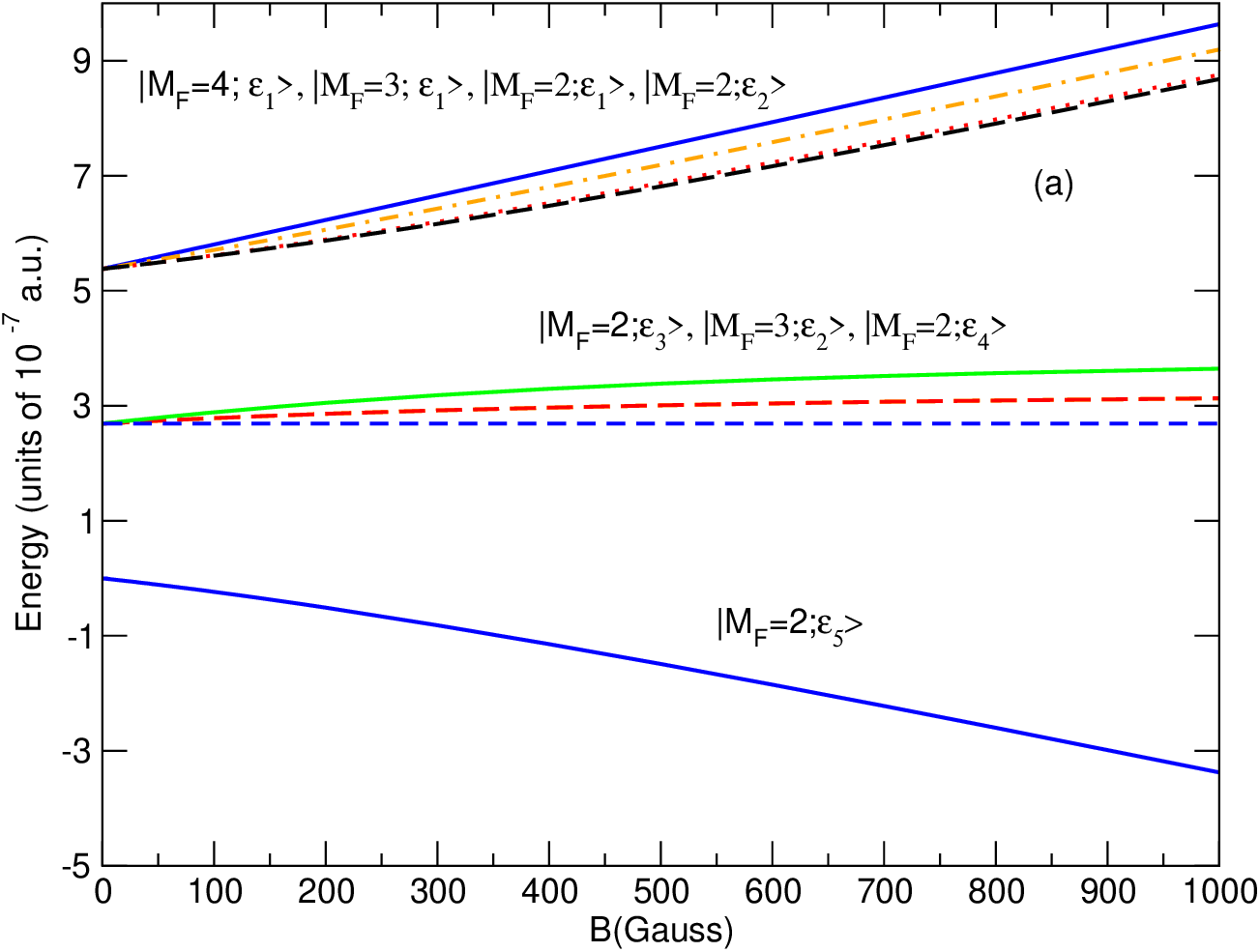}
\includegraphics[scale=.5,angle=0]{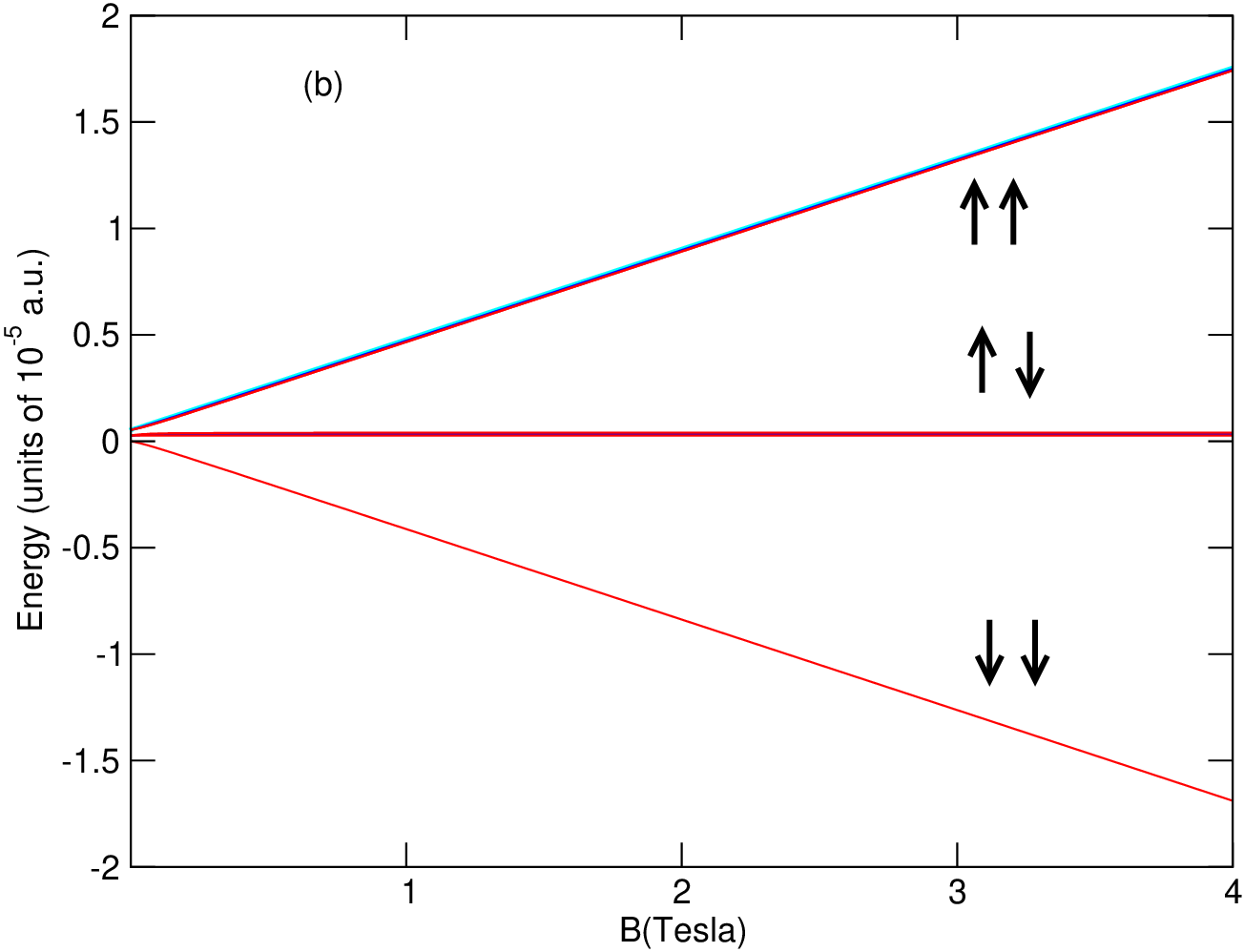}
\caption{\label{fig:1}  
 (a) Energy levels for a pair of sodium atoms in a
magnetic field. The levels correspond to
the states
itemized in Table II. In the figure they are grouped
in order of decreasing energy.
(b) In the large $ B$ field limit the arrows represent the
total electronic, azimuthal, spin angular momentum of
the corresponding levels shown. }
\end{figure}

\section{Close coupling expansion}
In the Pauli approximation, the magnetic Breit interaction between the
two valence electrons  is given by \cite{bs57}
\beq
\alpha^2 \Biggl [ - {8 \pi \over 3} 
 {\bf S}_{1} \cdot {\bf S}_{2} \, \delta^{3} ({\bf r}_{12}) 
+ {1 \over r_{12}^{3} } \Bigl \{ {\bf S}_{1} \cdot {\bf S}_{2} 
 - 3{ {\bf S}_{1} \cdot {\bf r}_{12} \, {\bf S}_{2} \cdot {\bf r}_{12} \over  r_{12}^{2} } 
  \Bigl \} \Biggr ]
\label{xxx1}
\eeq
where $ \alpha  $ is the fine structure constant,
$ {\bf S}_{i}$ the spin of electron $i$ and $ {\bf r}_{ij} $ the displacement
vector for the two electrons.  In order to include the magnetic
interactions in the scattering equations
we substitute $ {\bf r}_{ij}  \rightarrow  {\bf R} $, where $ {\bf R} $ is
the inter-nuclear vector of the two atoms. This approximation is valid
at large inter-nuclear separations and we replace expression (\ref{xxx1}) 
by the model interaction
\beq
H_{di-polar} \equiv
 { \alpha^2 \over R^{3} } \Biggl [ 
   {\bf S}_{1} \cdot {\bf S}_{2}   - 3 { ( {\bf S}_{1} \cdot {\bf R} )
({\bf S}_{2} \cdot { \bf R} ) \over { R^{2} } } \Biggr ]
\label{xxx2}
\eeq
where we have ignored the Fermi-contact term. 
We did not include the electron spin-nuclear spin and the
nuclear spin-nuclear spin interactions since
 they contribute to the interaction energy an amount that
is at least three orders of magnitude smaller 
than the interaction energy
obtained from Eq. (\ref{xxx2}).

Using standard Racah-algebra techniques, 
we re-express $ H_{di-polar} $ in terms of irreducible tensor operators,
thus \cite{weis78}
\beq
H_{di-polar} = v(R)  \sum_{q} (-1)^{q} {Y^{(2)}}_{q}(\theta \phi) S^{(2)}_{-q}
\label{xxx3}
\eeq
where $ Y^{(2)}_{q} $ are the components of the spherical harmonic of rank 2, $ S^{(2)} $
is a second rank tensor in the product space spanned by the states
 $ |S_{1} m_{1} S_{2} m_{2} > $, and
  $ v(R) \equiv - \sqrt{2 4 \pi \over 5}  {\alpha^2 \over  R^3} $ 
  expressed in atomic units.  

In the special of case of a null external magnetic field, both
 $ |FM_{F} f_{a} f_{a}>$ and $ |f_{a} m_{a} f_{b} m_{b}>$
basis vectors, itemized in Table I, diagonalize the total
Hamiltonian in the asymptotic region. We use the former set
to express 
the system wavefunction  by the close coupling expansion,
\beq
\Psi({\bf R},{\bf r}) = \sum F_{F M_{F} f_{a} f_{b} } ({\bf R})
| F M_{F} f_{a} f_{b} > 
\label{xxx4}
\eeq
where the sum is over all quantum numbers
itemized in Table I, and
 from which we obtain the coupled equations,
\beq
- {1 \over 2 \mu} {\nabla^{2}} F_{i} ({\bf R})
+ \sum_{j} V^{e}_{ij}(R) F_{j} ({\bf R}) +
 \sum_{j} V^{hf}_{ij} F_{j} ({\bf R}) +
 \sum_{j} V^{m}_{ij}({\bf R}) F_{j} ({\bf R}) = E F_{i} ({\bf R}).
\label{xxx5}
\eeq
In deriving Eq.(\ref{xxx5}) we have ignored non-adiabatic effects\cite{zyg03,wol83}, since
they are expected to provide a small correction in the sodium-sodium
system. 
The subscript on the scattering amplitude denotes the channel quantum numbers,
$ i \equiv ( {F M_{F} f_{a} f_{b} }) _{i} $, $ E$ is the total energy,
 and $  V^{e}_{ij}(R) $, $ V^{hf}_{ij}(R) $,
 $V^{m}_{ij}({\bf R})$ are multi-channel potentials that correspond to the 
 electrostatic, nuclear hyperfine, and di-polar magnetic interaction Hamiltonian
 respectively. The explicit electrostatic and hyperfine terms are
\beq
{V_{ij}}^{e}(R)=
\sum_{S I} (-1)^{f_{a}+f_{b}+f'_{a}+f'_{b}} \,
  \delta_{F,F'} \, \delta_{M_{F},M_{F'}} [F,S,I][f_{a},f_{a'},f_{b},f_{b'}]^{1/2}   
 \times \nonumber \\
\begin{Bmatrix}  1/2 & 3/2 & f_{a} \cr
                   1/2 & 3/2 & f_{b} \cr
                   S  &  I   & F    
\end{Bmatrix}
\begin{Bmatrix}  1/2 & 3/2 & f_{a'} \cr
                   1/2 & 3/2 & f_{b'} \cr
                   S  &  I   & F   
\end{Bmatrix}
\epsilon_{S}(R)
\label{xxx6}
\eeq
\beq
{V_{ij}}^{hf}= \delta_{ij} {A_{f} \over 2} \bigl [ f_{a}(f_{a}+1)+f_{b}(f_{b}+1) \bigr ]  
\label{xxx7}
\eeq 
where  $ A_{F}$ is the 
Fermi hyperfine constant for the ground state of sodium, 
$ \mu $ is the nuclear reduced mass\cite{zyg03,wol83}, and $\epsilon_{S}(R)$ are
the $ \rm Na_{2}$ ground state Born-Oppenheimer potentials, for the triplet
$S=1$, and singlet $S=0$ states respectively. In deriving
Eq. (\ref{xxx6}) we used the relation
\beq
< I M_{I} S M_{S} | F M_{F} f_{a} f_{b}> & =&
[F,f_{a},f_{b},S,I]^{1/2} (-1)^{F+M_{F}}
\times \nonumber \\
\begin{pmatrix}  S & F & I \cr M_{S} & -M_{F} & M_{I} \end{pmatrix}
\begin{Bmatrix}  1/2 & 3/2 & f_{a} \cr
                   1/2 & 3/2 & f_{b} \cr
                   S  &  I   & F        \end{Bmatrix}. 
\label{xxx8}
\eeq
We now derive an expression for the di-polar interaction.
Using Eq. (\ref{xxx3}) we have
\beq
V^{m}_{ij}({\bf R}) =
 v(R) \sum_{q} (-1)^{q}  Y^{(2)}_{q}(\theta \phi)
  <f_{a'} f_{b'} M_{F'} F' | S^{(2)}_{-q} |F M_{F} f_{a} f_{b}>
\label{xxx9} 
\eeq
but
\beq
<f_{a'} f_{b'} M_{F'} F' | S^{(2)}_{-q} |F M_{F} f_{a} f_{b}> = 
  \sum_{S M_{S}I M_{I}}  \sum_{S' M_{S'}I' M_{I'}} 
   (-1)^{F+F'+M_{F}+M_{F'}} \delta_{I,I'} \delta_{M_{I},M_{I'}}  \times \nonumber  \\ 
   {[ F,F',S,S',I,I',f_{a},f_{a'},f_{b},f_{b'}]}^{1/2} <SM_{s}| S^{(2)}_{-q} |S' M_{S'}> 
\begin{Bmatrix} S & F & I \cr M_{S} & -M_{F} & M_{I} \end{Bmatrix}
 \times  \nonumber \\
\begin{pmatrix} S' & F' & I \cr M_{S'} & -M_{F'} & M_{I}  \end{pmatrix}
\begin{Bmatrix} 1/2 & 3/2 & f_{a} \cr
                   1/2 & 3/2 & f_{b} \cr
                   S  &  I   & F   \end{Bmatrix}
\begin{Bmatrix}  1/2 & 3/2 & f_{a'} \cr
                   1/2 & 3/2 & f_{b'} \cr
                   S'  &  I'  & F'   \end{Bmatrix} 
\label{xxx10}
\eeq
and since 
\beq
<SM_{s}| S^{(2)}_{-q} |S' M_{S'}> & =& (-1)^{S-M_{S}}                  
\begin{pmatrix} 
S & 2 & S' \cr -M_{S} & -q & M_{S'} \end{pmatrix} <S||S^{(2)}||S'>=
  \nonumber \\
& &\delta_{S,S'} \delta_{S,1} \sqrt{{15 \over 12}}
(-1)^{M_{S}+1} \begin{pmatrix} 
S & 2 & S \cr -M_{S} & -q & M_{S'} \end{pmatrix}
\label{xxx11}
\eeq
we get
\beq
<f_{a'} f_{b'} M_{F'} F' | S^{(2)}_{-q} |F M_{F} f_{a} f_{b}>  &= &
  \sum_{M_{S} M_{S'}} \sum_{I M_{I}}  
   (-1)^{F+F'+M_{F}+M_{F'} + M_{S}+1} 
    \times \nonumber \\  {[F,F',f_{a},f_{a'},f_{b},f_{b'}]}^{1/2} \sqrt{{ 15 \over 12}}& &
       \begin{pmatrix} 
1 & 2 & 1 \cr -M_{S} & -q & M_{S'} \end{pmatrix}    
\begin{pmatrix} 1 & F & I \cr M_{S} & -M_{F} & M_{I} \end{pmatrix}
 \times \nonumber \\
\begin{pmatrix} 1 & F' & I \cr M_{S'} & -M_{F'} & M_{I'} \end{pmatrix}
 & & \begin{Bmatrix}  1/2 & 3/2 & f_{a} \cr
                   1/2 & 3/2 & f_{b} \cr
                   1  &  I   & F    \end{Bmatrix}
\begin{Bmatrix}  1/2 & 3/2 & f_{a'} \cr
                   1/2 & 3/2 & f_{b'} \cr
                   1  &  I   & F'    \end{Bmatrix}. 
\label{xxx12}
\eeq
Therefore,
\beq
V^{m}_{ij}({\bf R}) &=&
  \sum_{q} (-1)^{q}  Y^{(2)}_{q}(\theta \phi) v_{ij}(q,R) \nonumber \\ 
 v_{ij}(q,R)& \equiv & v(R)  \sum_{M_{S} M_{S'}} \sum_{I M_{I}}  
   (-1)^{F+F'+M_{F}+M_{F'} + M_{S}+1}   
     {[F,F',f_{a},f_{a'},f_{b},f_{b'}]}^{1/2} \times  \nonumber \\
    \begin{pmatrix} 1 & 2 & 1 \cr -M_{S} & -q & M_{S'} \end{pmatrix} 
& & \begin{pmatrix} 1 & F & I \cr M_{S} & -M_{F} & M_{I} \end{pmatrix}
\begin{pmatrix} 1 & F' & I \cr M_{S'} & -M_{F'} & M_{I} \end{pmatrix}
                                                \times  \nonumber \\
& & \sqrt{{ 45 \over 4}}   \begin{Bmatrix}  1/2 & 3/2 & f_{a} \cr
                   1/2 & 3/2 & f_{b} \cr
                   1  &  I   & F   \end{Bmatrix}
\begin{Bmatrix}  1/2 & 3/2 & f_{a'} \cr
                   1/2 & 3/2 & f_{b'} \cr
                   1  &  I   & F'   \end{Bmatrix}.
\label{xxx13}
\eeq

We express amplitude (\ref{xxx4}) by a partial wave expansion in
spherical harmonics,
\beq
F_{i}({\bf R}) = \sum_{l m} 
 { F_{i}(lm,R) \over R} Y_{lm}(\theta \phi).
 \label{yyy1}
 \eeq
Though $V_{ij}^{e}, \, V_{ij}^{hf}$ are isotropic in the nuclear orientation
$ V_{ij}^{m}$, according to Eq. (\ref{xxx13}),
 is not and the partial wave expansion does not lead to
 radial equations that are diagonal in the nuclear angular momentum $l$ and $m$.
Inserting (\ref{yyy1}) into 
Eq. (\ref{xxx5}) we obtain,
\beq
- {1 \over 2 \mu}\bigl ( {d^2 \over dR^2} -{l(l+1) \over R^{2}} \bigr ) F_{i}(lm,R)
+ \sum_{j} V^{e}_{ij}(R) F_{j}(lm,R) +
 \sum_{j} V^{hf}_{ij} F_{j}(lm,R) + \nonumber \\
 \sum_{j} \sum_{l',m'}u_{ij}(lm,l'm',R)F_{j}(l'm',R)   = E F_{i}(lm,R)
\label{yyy2}
\eeq
where,
\beq
u_{ij}(lm,l'm',R) &=& ({[l,l',2] \over  4 \pi})^{1/2} \sum_{q} v_{ij}(q,R)
  (-1)^{q+m} \begin{pmatrix}l & 2 & l' \cr -m & q & m' \end{pmatrix} 
 \begin{pmatrix} l& 2 & l' \cr 0 & 0 & 0 \end{pmatrix}
\label{yyy3}
\eeq
According to Eq. (\ref{xxx13}) $ q=M_{F}-M_{F'}\equiv \Delta M_{F} $,
and from Eq. (\ref{yyy3}) $q=m-m' \equiv \Delta m$. Therefore we
obtain the selection rule $ \Delta m = \Delta M_{F} $ and from
the selection rules for the $ 3j $ symbols in Eq. (\ref{yyy2}) we
require $ l-l' \equiv \Delta l =0,2$ , and $ l=l'\neq 0$.

\section{Scattering formalism}
It is useful to re-express the coupled radial equations
in the form
\beq
- {1 \over 2 \mu}\bigl ( {d^2 \over dR^2} -
{{\underbar l}({\underbar l}+1) \over R^{2}} \bigr ) {\underbar F}(R)
+ {\underbar V}^{e}(R)  {\underbar F}(R)
+  {\underbar V}^{hf} {\underbar F}(R) + {\underbar V}^{m}(R){\underbar F}(R)
  &=& \nonumber \\ 
  E {\underbar F}(R). & &
\label{zzz1}
\eeq
In the notation introduced above $ {\underbar F}(R) $ is a square matrix
whose columns contain the independent solution vectors
 to the coupled equations (\ref{yyy2}). The row
and column indices  for matrix $  {\underbar F}(R) $ itemize
both the internal and orbital angular momentum quantum numbers.
A  given value of index $i$, identifies
 the set  $ i \equiv
 \{(F M_{F} f_{a} f_{b} )_{i} l_{i} m_{i} \} $
 where $l_{i}$, $m_{i}$ are the total
and azimuthal quantum numbers for channel $i$.
 At a given
collision energy, we are allowed to truncate the partial wave expansion 
(\ref{yyy1}) at some maximum value $l_{max} $ and  matrix
 $ {\underbar F}(R) $ is a finite $n$-dimensional square matrix where
 $n= n_{i} \prod_{l=0}^{l_{max}}(2 l+1)$, 
 and $n_{i}$ is the dimension of the internal Hilbert space. 

Matrices 
$ {\underbar V}^{e}(R) $, $ {\underbar V}^{hf}$,
correspond to the electrostatic and hyperfine Hamiltonian respectively,
 and they are diagonal with respect
to the angular momentum quantum numbers. The matrix 
${\underbar l}$ is diagonal and contains 
the channel angular momenta along the diagonal.
However, $  {\underbar V}^{m}(R)$,  whose components
 are $  {\underbar V}^{m}_{ij}(R)=u_{ij}(l_{i}m_{i};l_{j}m_{j};R) $,
 is not diagonal.

In the limit $ R \rightarrow \infty$  we require that 
\beq
{\underbar F}_{ij}(R) \rightarrow {1 \over \sqrt{k_{i}} } 
\Biggl \{  \delta_{ij} sin(k_{i} R - l_{i} {\pi \over 2}) + 
{\underbar K}_{ij}cos(k_{i} R - l_{i} {\pi \over 2} ) \Biggr \}
\label{zzz2}
\eeq
where  ${\underbar K}_{ij} $ are the elements of the $ K$ matrix.

We introduce the amplitude 
 $ {\underbar G}(R) \equiv {\underbar F}(R) \, {\underbar C} $
 where ${\underbar C}$ is a constant matrix chosen so that in the limit 
 $ R\rightarrow \infty $
 \beq
 {\underbar G}_{ij}(R) = { 1 \over k_{i}^{1/2} } 
  \Biggl \{  \delta_{ij} \exp(-i(k_{i} R - l_{i} {\pi \over 2})) - 
  {\underbar S}_{ij} \exp(i(k_{i} R - l_{i} {\pi \over 2})) \Biggr \}
\label{zzz3} 
\eeq
where the radial S-matrix $ {\underbar S} $ is,
\beq
{\underbar S} = ({\underbar I} - i {\underbar K}) 
 ({\underbar I} + i {\underbar K})^{-1}
\label{zzz4}
\eeq
We construct a reduced multichannel amplitude, ${\underbar G}_{[ij]}({\bf R}) $, 
where the notation $[ij]$ implies that the indices denote
the quantum numbers of the internal
states only. We define
\beq
{\underbar G}_{[ij]}({\bf R})= \sum_{l_{i}m_{i}} \sum_{l_{j}m_{j}}
Y_{l_{i}m_{i}}(\theta \phi) Y^{*}_{l_{j}m_{j}}(\theta_{i} \phi_{i}) 
  { 2 \pi i^{l_{j}+ 1} \over k_{j}^{1/2} } 
{ {\underbar G}_{ij}(R) \over R} 
  \label{zzz5}
  \eeq
and find that
 in the asymptotic limit $ R \rightarrow \infty $ 
  \beq
  {\underbar G}_{[ij]}({\bf R}) \rightarrow  \delta_{[ij]} 
  \exp(i {\bf K}_{i} \cdot {\bf R}) + f_{[ij]} (\theta \phi; \theta_{i} \phi_{i}) 
  { \exp(i k_{i} R) \over R }
  \label{zzz6} 
  \eeq 
where we have used Eq. (\ref{zzz3}), and 
$  f_{[ij]} (\theta \phi; \theta_{i} \phi_{i}) $ is the scattering amplitude for the system
to undergo a transition from an initial internal state $ j$ into
an internal state $i$
 and into solid angle 
$ d \theta (sin\theta) d \phi $ following an initial approach along the incident wave
vector $ {\bf K}_{i} $ with polar angles $ \theta_{i} \phi_{i} $. Comparing
expression (\ref{zzz5}) and (\ref{zzz6}) we find that
\beq
f_{[ij]} (\theta \phi; \theta_{i} \phi_{i}) \equiv
  \sum_{l_{i} m_{i}} \sum_{l_{j} m_{j}}
Y_{l_{i}m_{i}}(\theta \phi) Y^{*}_{l_{j}m_{j}}(\theta_{i} \phi_{i})
 { 2 \pi i^{l_{j}+1}  \over k_{i}^{1/2} k_{j}^{1/2} }
  ( \delta_{ij} - {\underline S}_{ij}).
 \label{zzz7}
 \eeq 

Though $ {\underbar G}_{[ij]}({\bf R}) $ has the desired asymptotic behavior for 
scattering solutions it does not posses the symmetry required
by the Pauli principle. Because the sodium nuclei are identical fermions, the total
wavefunction must be odd under their interchange. Let $ P^{N}_{12} $ be
the nuclear permutation operator, then\cite{zyg94}
\beq
P^{N}_{12} | F M f_{a} f_{b} > = (-1)^{F+f_{a}+f_{b}+1} | F M f_{b} f_{a} >. 
\label{zzz8}
\eeq
We introduce a shorthand notation for the channel indices that label
the matrix $ {\underbar G}({\bf R})$; if $ i={F M f_{a} f_{b} } $ then
$ {\tilde i} \equiv {F M f_{b} f_{a} } $. In this notation the above
relation is written $ P^{N}_{12} | i >=(-1)^{F+f_{a}+f_{b}+1} |{\tilde i}> $.
If the system is
  initially prepared in state $j$, and is given by $ |\Psi> = 
\sum_{i} G_{[i j]}({\bf R}) |i> $ , then we 
require that $ G_{[ij]}(-{\bf R})=
(-1)^{F+f_{a}+f_{b}} G_{[{\tilde i} j]}({\bf R})$. 

Using this notation we replace (\ref{zzz5}) with,
\beq
 G_{[i j]}({\bf R}) \equiv 
\sum_{l_{i} m_{i}} \sum_{l_{j} m_{j}}
Y_{l_{i}m_{i}}(\theta \phi) Y^{*}_{l_{j}m_{j}}(\theta_{i} \phi_{i})
 { 2 \pi i^{l_{i}+1}  \over k_{j}^{1/2} }
   \times \nonumber \\
  \Biggl \{ {G_{ij}(R) \over R} +(-1)^{F+l_{i}+f_{a}+f_{b}} 
  { G_{{\tilde i} j}(R) \over R}  \Biggr \}
\label{zzz11}
\eeq
which has the desired symmetry.

In the asymptotic limit we get,
\beq
G_{[ij]}({\bf R}) & \rightarrow &
  \delta_{[i j]} 
 \exp(i {\bf K}_{i} \cdot {\bf R}) + 
 (-1)^{F+f_{a}+f_{b}} \delta_{[{\tilde i}j]} 
\exp(-i {\bf K}_{i} \cdot {\bf R})
 + \nonumber \\
  \Bigl [  f_{[ij]}(\theta \phi ; \theta_{i} \phi_{i}) 
   & + & (-1)^{F+f_{a}+f_{b}} 
    f_{[{\tilde i} j]} (-\theta + \pi \, \phi + \pi ;  \theta_{i} \phi_{i} ) \Bigr ]
   { \exp ( i k_{i} R) \over R } \label{zzz12}
\eeq
where we have used $ k_{i}=k_{{\tilde i}} $.
The cross section for a system in an internal state $ |j>=|F'M'f_{a'}f_{b'}>$ to
undergo a transition into state  $ |i>=|F M f_{a} f_{b}>$ is
\beq
\sigma( j \rightarrow i)  &=& 
{ v_{i} \over v_{j} } { 1 \over 2} \, {1 \over 4 \pi} \times \nonumber \\
& & \int d {\hat {\bf \Omega}} \int d {\hat {\bf \Omega_{i}}} | 
 f_{[ij]}(\theta \phi ; \theta_{i} \phi_{i}) 
 + (-1)^{F+f_{a}+f_{b}} f_{{[\tilde i}j]}
 (-\theta + \pi \, \phi + \pi ;  \theta_{i} \phi_{i} ) |^{2}
\label{zzz12a}
\eeq
where we integrate over all scattering angles and average over all directions
of the incident wave. $ v_{j} $ is the velocity in the incoming
channel and $  v_{i} $ the final channel velocity. 
We have included a factor of $ {1 \over 2}$
in order to insure that the incoming flux is normalized to unity. Using expression
(\ref{zzz11}) we can re-write Eq. (\ref{zzz12}), 
\beq
\sigma( j \rightarrow i ) & = & 
 {  \pi \over 2 k^{2}_{j} } 
  \sum_{l_{i} m_{i}} \sum_{l_{j} m_{j}} 
| T_{[ij]}(l_{i}m_{i};l_{j}m_{j})  
 + (-1)^{F +f_{a}+f_{b}+l_{i} } T_{[{\tilde i}j]}(l_{i}m_{i};l_{j}m_{j})|^{2}
 \nonumber \\
T_{[ij]}(l_{i}m_{i};l_{j}m_{j}) &\equiv&  \delta_{[ij]}-S_{[ij]}
\label{zzz13}
 \eeq

 We use Eq. (\ref{zzz13}) to calculate the total inelastic
 transition cross section in the case for zero, or small, magnetic
 field intensities. The initial state corresponds to the
 maximal extended state $ |F=4 \, M=4 \, f_{a}=2\, f_{b}=2 > $ and at
 low energies only incident s-waves contribute.  According
 to the dipolar selection rules the exit channels are
 d-waves, and we obtain a simple expression for the total
 inelastic cross section
\beq
 \sigma_{T} =  \sum_{FM_{F}f_{a} f_{b}} \,  \sum_{m_{i}=-2}^{m_{i}=2}
  {2 \pi \over k^{2}} \, |{\tilde T}_{FM_{F}f_{a}f_{b}} (m_{i})|^2
 \label{aaa1}
\eeq
where,
 $ {{\tilde T}}_{FM_{F}f_{a}f_{b}}(m_{i}) \equiv
  T_{[ij]}(l_{i}=2,m_{i};l_{j}=0,m_{j}=0)$  for
 $ i={FM_{F}f_{a}f_{b}} $
 and $ j={F=4 \,M_{F}=4,f_{a}=2 \, f_{b}=2 }$, and $ k=k_{j}$.
 In deriving Eq. (\ref{aaa1}) we used the fact
 $ T_{[{\tilde i} j]} = (-1)^{F+f_{a}+f_{b}} T_{[i j]}  $.

 For a large magnetic field, such that the Zeeman
splitting is much greater than the hyperfine interaction,
we construct close coupling equations by  using the basis
vectors $ |  S M_{S} I M_{I} > $ in expansion Eq. (\ref{xxx4}).
We obtain an equation  analogous to  Eq. (\ref{zzz1}) except
that $ {\underline V}^{hf} $ is replaced by an expression
that describes the Zeeman interaction with the external field.
In addition, the electrostatic and dipolar interaction 
matrices are replaced by 
 $ {\underline {\tilde V}}^{e}(R) $
and $ {\underline {\tilde V}}^{m}(R) $ respectively.
They are related by the unitary transformation
$ {\underline {\tilde V}}^{e}(R) =
 {\underline U}  {\underline  V}^{e}(R) {\underline U}^{-1} $,
$ {\underline {\tilde V}}^{m}(R) =
 {\underline U}  {\underline  V}^{m}(R) {\underline U}^{-1} $,
where
$U_{ij} = <F M_{F} f_{a} f_{b} l_{j} m_{j}  |l_{i} m_{i}  S M_{S} I M_{I} >$.

 Because the
states $ | S M_{S} I M_{I}  > $ are eigenstates of the nuclear
interchange operator $ P_{12}^{N} $, i.e.
\beq
 P_{12}^{N} | S M_{S} I M_{I}  > = 
(-1)^{S+I+1} |S M_{S} I M_{I}  >
\label{aaa2}
\eeq
we obtain
\beq
\sigma( j \rightarrow i ) & = &
 {  \pi \over 2 k^{2}_{j} }
  \sum_{l_{i} m_{i}} \sum_{l_{j} m_{j}}
| T_{[ij]}(l_{i}m_{i};l_{j}m_{j})(1 +
  (-1)^{I+S+l_{i}} ) |^{2}
\label{aaa3}
 \eeq
where the channel indices now itemize the states in Table 1
under the $ | S M_{S} I M_{I} > $ representation.

If the magnetic interaction energy is of the same order as the
hyperfine energy, neither the $ | F M_{F} f_{a} f_{b}> $ nor
the $ |  S M_{S} I M_{I}> $ representations constitute a valid 
asymptotic basis since off-diagonal elements persist at large
inter-nuclear separations. Instead, we choose a linear combination
of these states that diagonalize the asymptotic Hamiltonian,
\beq
H^{hf} + H^{Z} \nonumber  \\
H^{Z} = 2 {\mu_{B} }  {\bf B} \cdot {\bf S}
\label{aaa4}
\eeq
where $ H^{hf}$ is the hyperfine interaction, $ {\bf B}$
is the external magnetic field whose orientation defines our
lab quantization axis, $ {\bf S}$ is the total electronic spin
for the atom-atom system and $ \mu_{B} $ is the Bohr magneton.
We ignored the magnetic-nuclear
 term  since it provides a considerable 
smaller contribution to the total magnetic interaction energy 
than that given by Eq. (\ref{aaa4}).     
The diagonalization procedure can be carried out numerically,
and  in Table II we itemize those states which contribute
to dipolar loss from the incident, extended state, channel.
Good quantum numbers for these states include the total
azimuthal quantum number $ M_{F}$, the nuclear interchange
parity, and the energy eigenvalues $ e_{i} $ itemized in
Table II. The extended state is odd under nuclear
interchange i.e., $ P^{N}_{12} | M_{F}=4; \, p=1 \, e_{1}> =
 (-1)^{p}  \, | M_{F}=4; \, p=1 \, e_{1}> $ where we have used the notation
described in Table II.
\begin{table}
 \caption{Quantum numbers associated with  the states
$ |M_{F}\, p \, \epsilon_{i}> $ that diagonalize
 the asymptotic Hamiltonian, Eq. (\ref{aaa3}).
We itemize only those states
whose parity, under nuclear interchange, is odd.
The parameter $ \zeta \equiv { 4 \mu_{B} B \over \Delta F} $
where $B$ is the magnetic field strength and $ \Delta_{F}$
the energy defect between the $ F=2$ and $F=1$ hyperfine
levels of Sodium. The last column itemizes the $ B \rightarrow 0 $
limit of the states expressed in the $|F M_{F} (f_{a} f_{b})> $ representation.}
\begin{ruledtabular}
\begin{tabular}{c|c|c|c}
 $\epsilon$  & $ M_{F} $ &  Energy  & $| M_{F} \, p \, \epsilon  > $  \hfill \\
\hline
 $\epsilon_{1} $& 4 & 2 $ \Delta_{F} (1+ {\zeta \over 4}) $ & $ |4422>$  \\
\hline
 $\epsilon_{1} $ & 3 &  $ { \Delta_{F} \over 4}(6+\zeta +\sqrt{4+2 \zeta +\zeta^{2}} )$
 & $ |4322>$  \\
 $\epsilon_{2} $ & 3 &  $ { \Delta_{F} \over 4}(6+\zeta -\sqrt{4+2 \zeta +\zeta^{2}} ) $ &
 $|33(12)>$ \\
\hline
$\epsilon_{1} $ & 2 & $  { \Delta_{F} \over 2} ( 2 + \sqrt{4 +2 \zeta +\zeta^2} ) $ &
$ {2 \over \sqrt{7} } |4222>  -\sqrt{3 \over 7}|2222> $ \\
$\epsilon_{2} $ & 2 & $  { \Delta_{F} \over 4} (6+\zeta+\sqrt{4+\zeta^2} ) $ &
$ -\sqrt{3 \over 7} |4222>  -{2 \over \sqrt{7} }|2222> $ \\
$\epsilon_{3} $ & 2 & $  { \Delta_{F} \over 4} (6+\zeta-\sqrt{4+\zeta^2} ) $ &
 $ {1 \over \sqrt{3} }|32(12)> -\sqrt{2 \over 3} |22(12)> $ \\
$\epsilon_{4} $ & 2 & $    \Delta_{F} $&
$ -\sqrt{2 \over 3}|32(12)> -{ 1 \over \sqrt{3}} |22(12)>   $ \\
$\epsilon_{5} $  & 2 & $  { \Delta_{F} \over 2} ( 2 - \sqrt{4 +2 \zeta +\zeta^2} ) $ & $ |2211> $  \\
\end{tabular}
\end{ruledtabular}
\end{table}

 Only states of odd parity are
allowed as exit channels and these states are listed in Table II.
Invoking the procedure discussed above, we obtain for the
total dipolar loss cross section
\beq
 \sigma_{T} =  \sum_{j} \,  \sum_{m_{i}=-2}^{m_{i}=2}
  {2 \pi \over k^{2}} \, |{\tilde T}_{j} (m_{i})|^2
 \label{aaa5}
\eeq  
where the sum over index $j$ denotes the channels itemized
in Table II.

The rate coefficient for dipolar relaxation is given
by the expression
\beq
k_{T} = \sqrt{  {8 k T \over \pi \mu }  }
 ({ 1 \over kT })^2 \int_{0}^{\infty} dE \, E \, \sigma_{T}(E) 
\exp(-{E \over kT})
\label{aaa6}
\eeq
where $ \mu $ is the reduced mass of the $ \rm ^{23} Na_{2} $
system and $ \sigma_{T}(E) $ the total 
inelastic cross section expressed as a function of
collision energy.

\section{Results and discussion}

In Fig. 2 we plot the total rate coefficient(solid line), in the
 $T \rightarrow 0$ limit, as
a function of magnetic field strength. 
 In Fig. 2 we notice
that the total relaxation rate is nearly constant for field strengths
up to about 100 Gauss(G). In the range $ \rm{100 \, G < B < 400 \, G}$
 the rates exhibit significant structure.
  For larger values of $ \rm B$ the total
rate diminishes in a  monotonic manner. 
In figure 2 we also plot, shown by the dashed line,
results obtained using the approximation where
the hyperfine states $ |F M_{F} f_{a} f_{b}>$ define the
asymptotic basis states and  the asymptotic off-diagonal terms,
due to magnetic interactions, are neglected. For small
$\rm B$ the approximation gives excellent agreement, for the total
dipolar loss rate, with the results obtained using the appropriate
$ |M_{F} \, p\, \epsilon > $ basis. However, for $ {\rm B > 60 \, G}$
this approximation considerably overestimates the total rate.

\begin{figure}
\includegraphics[scale=.5,angle=0]{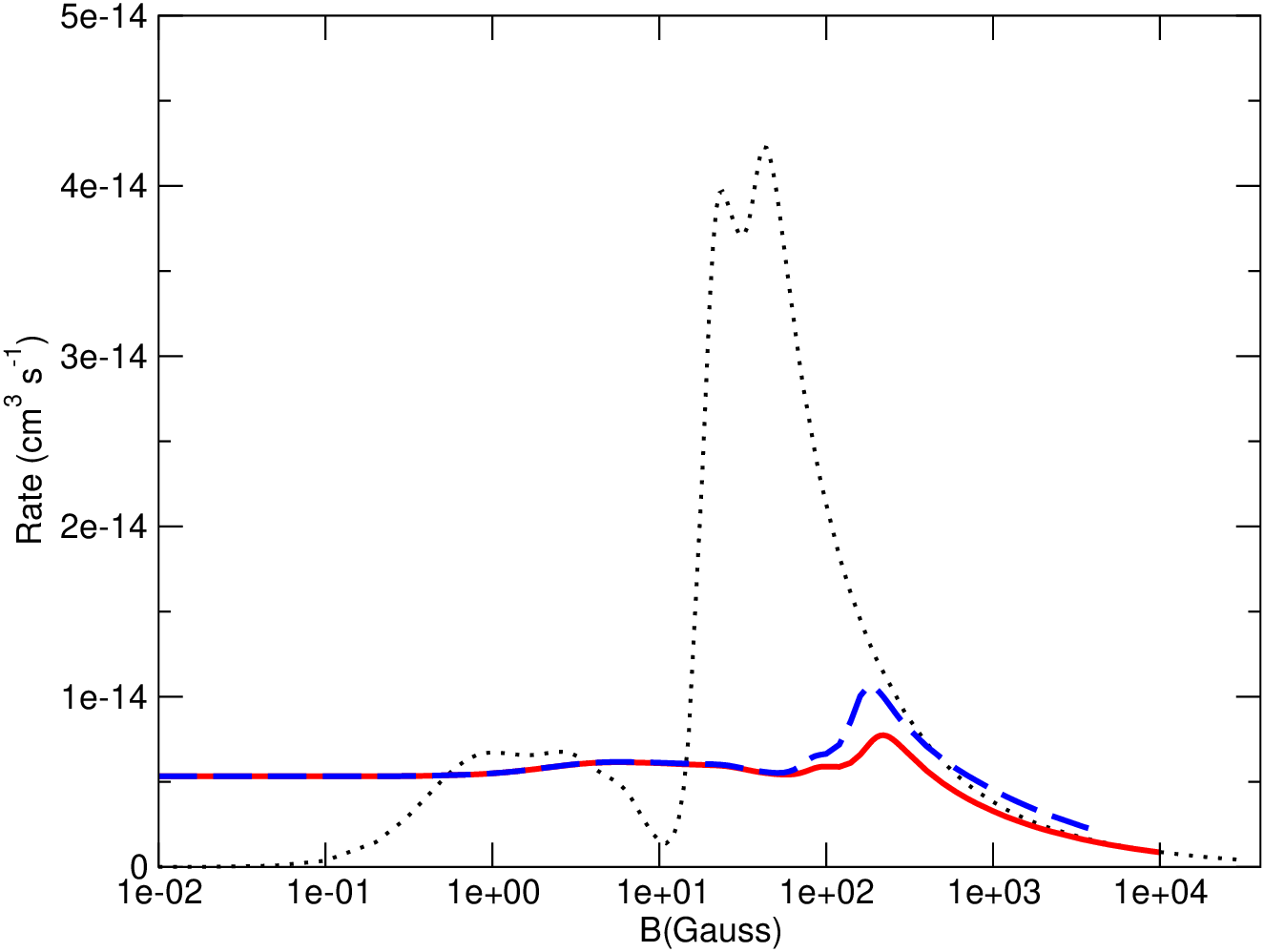}
\caption{\label{fig:2}   Total dipolar relaxation rate (heavy solid line) as a function
of external magnetic field strength. The dashed lines corresponds
to the approximation where the hyperfine basis $ |F M_{F} f_{a} f_{b}>$
is used for the asymptotic channel states. The dotted line
corresponds to the case where the $ |S M_{S} I M_{I}>$ basis
is used and the hyperfine interaction is ignored. }
\end{figure}

The dotted line in the figure presents the results of
the calculation when we used the $ | S M_{S} I M_{I}> $
representation to define the asymptotic channel basis, and
where we ignored the hyperfine Hamiltonian. For small
 B this approximation is poor because it neglects
the dominant hyperfine interaction. In the $ {\rm B \rightarrow 0}$ limit,
rates obtained in this approximation vanish due
to the nature of the anisotropic dipolar interaction. According
 to the selection rules discussed in the
previous sections,
a pair of atoms approaching as an s-wave must exit
as a d-wave, and therefore,  the exit channel must be exothermic
with respect to the entrance channel for the collision to proceed
in the zero temperature limit.
 At $ \rm B \rightarrow 0$ a finite energy defect is generated by
the hyperfine interaction and, if hyperfine effects
are ignored, the dipolar rate vanishes in the
$ \rm T \rightarrow 0$, 
limit. This effect is clearly evident in Fig. 2. As 
$ \rm B \rightarrow \infty $ the neglect of the hyperfine
interaction is justified and, in that case, we expect that
the rate obtained using the $ | S M_{S} I M_{I}> $ 
basis to be a good approximation.
In Fig. 2 we note that for ${\rm B> 1 \, T}$ the  dotted line
merges with the solid line and illustrates the validity
of that approximation at large
field strengths.

To understand the nature of the observed structures in the
total rates we neglect the fine structure and study the
collision dynamics in the $| S M_{S} I M_{I}> $ basis and
show the results in Fig. 3. 
 In that figure, the dashed line denotes the partial rate into the state
where both atoms flip their total electronic spin, whereas
the solid line corresponds to the case where only one
atom flips its spin. 
We first consider the kinematics of the latter case.

In the $ | S M_{S} I M_{I}>$ basis the Zeeman energy splitting
between the $ | S=1; M_{S}=1; I=3; M_{I}> $ and the
  $ | S=1; M_{S}=0; I=3; M_{I}> $ exit channel is given by
\beq
\Delta E= |2 \mu_{B} {\rm B} | + {k^{2} \over 2 \mu}
\label{uuu1}
\eeq
where $k$ is the wavenumber that corresponds to the kinetic
energy of the system in the entrance channel and $\rm B$ is the
absolute value of the magnetic field that is parallel to the
laboratory z-axis. In the  $ k \rightarrow 0$ limit
the exit channel is a d-wave and, using
the potential for the $ ^{3} \Sigma_{u}$ state of
 $\rm ^{23}Na_{2}$ system tabulated by Samuelis et al. \cite{sam00}, we find
that the $l=2$ centrifugal barrier has a height 
$ \Delta E(l=2)=1.6585 \times 10^{-8} \, a.u. $ We equate the
Zeeman splitting with the barrier 
height and find that the critical magnetic field strength 
required so enough kinetic energy is available in the exit
channel
to overcome the barrier 
has the
value $ \rm B_{c} = 39.0 \, \rm G $.
\begin{figure}
\includegraphics[scale=.5,angle=0]{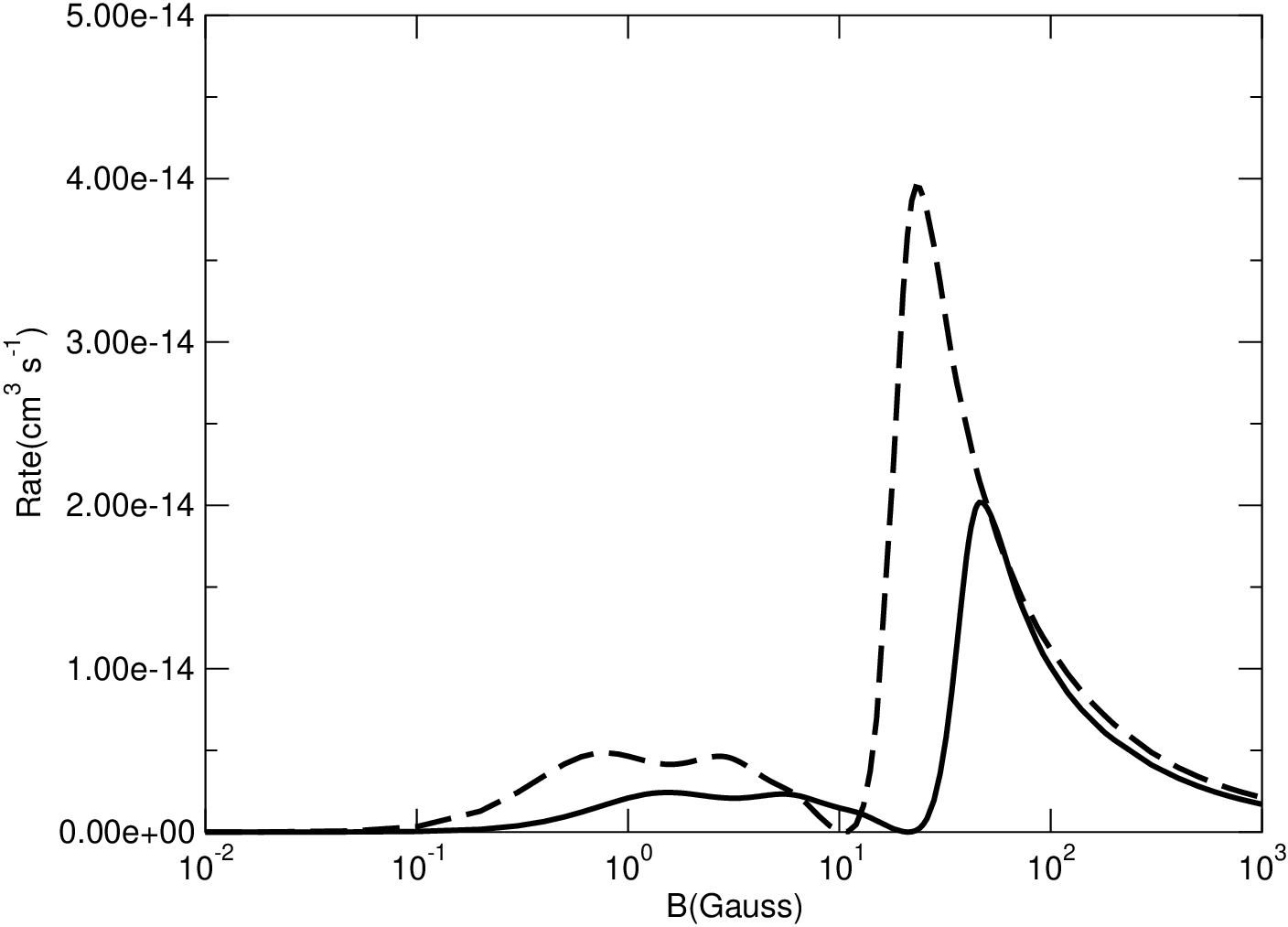}%
\caption{\label{fig:3} 
  Resonance/threshold  structures observed in the dipolar loss rates
that were calculated ignoring hyperfine effects and in the
$|S M_{S} I M_{I}>$ basis.
The dashed line corresponds to dipolar loss involving total
electronic spin flips for a single atom, whereas the solid
line corresponds to spin flips involving both atoms.
   }
\end{figure}

 According to Fig. 3, at this field strength
the relaxation rate is rapidly increasing, as $ \rm B$ increases,
 but it is in a region to
the left of its maximum which occurs at $ \rm B_{max} = 46.0 \,\rm G $.
Therefore, the pronounced structure seen in this rate cannot
be solely attributed to a threshold effect. Indeed we found  a
resonance in the $ l=2 $ partial wave that is due to the existence
of a virtual state at $ E_{v}= 1.85 \times 10^{-8} \,  a.u.$ 
Using Eq. (\ref{uuu1})
to convert $E_{v}$ into a field strength, we find
$ \rm B^{*} = 43.5 \, \rm G$, a value that is about midway between
 $\rm B_{c}$ and $ \rm B_{max}$.
For transitions into the state $ | S=1; M_{S}=-1; I=3; M_{I}> $,
whose rate coefficient is shown by the dashed line in Fig 4,
B
we evaluate $ \rm B^{*} = { E_{v} / 4 \mu_{B} } $ and 
get $ \rm B^{*}=21.8 \, \rm G $. This value is close to 
$ \rm B_{max}=23.0 \, \rm G$ seen in the figure.
 These observations
strongly suggest that the structure, evident in the
in Fig. 3, is a consequence of shape
resonance phenomena \cite{zyg02,laue02}. This conclusion is strengthened by
studies, discussed below, of this collision process at higher temperatures.

\begin{figure}
\includegraphics[scale=.5,angle=0]{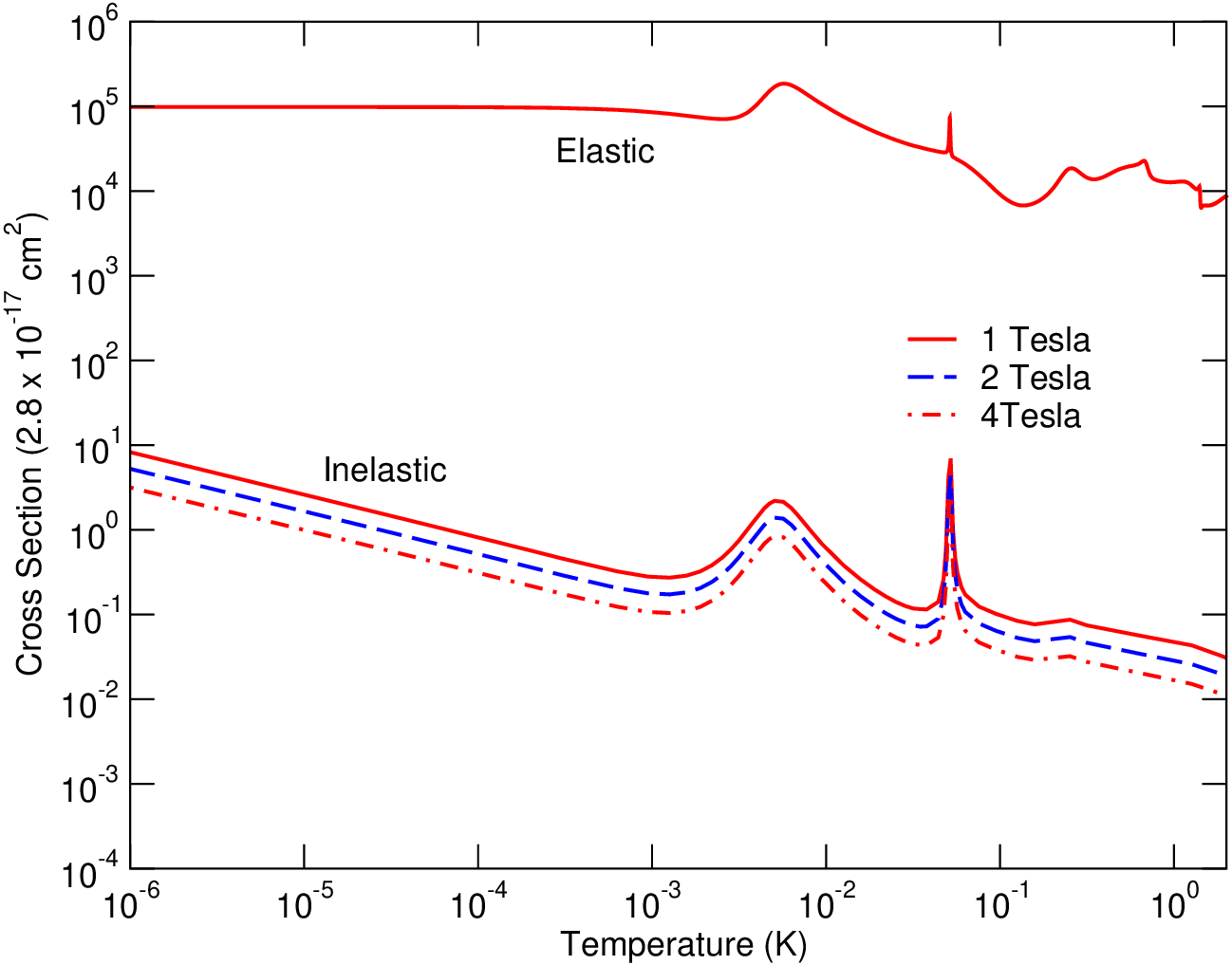}%
\caption{Elastic and inelastic cross sections in collisions of
${\rm Na}(F=2 \, M_{F}=2)+{\rm Na}(F=2 \, M_{F}=2).$
The collision energy is expressed in units of Kelvin. \label{fig:4}}
\end{figure}

Our rate in the $ \rm B \rightarrow 0$ limit is in harmony with that
that reported in a previous study \cite{ver96}, but several times
larger from that predicted in 
an earlier study \cite{ver91}. In Refs.\cite{ver96,ver91}
the authors used the 
$ |f_{a} m_{a} f_{b} m_{b}>$ basis to calculate the rates for
$ B \neq 0$. We have shown here, that this approximation 
overestimates the relaxation rates for $\rm B > 60 \, G$. The largest
uncertainty is probably associated with the choice for molecular
potentials of the ground $\rm ^{23} Na $ system. We adopted the
most recent, and accurate, potentials tabulated
by Samuelis et al. \cite{sam00}. 

In Fig. 5 we present the results
of our calculation for both the elastic and inelastic, dipolar relaxation,
cross sections in collisions of spin polarized sodium atoms. The 
collision energies considered range from ultra-cold to 2 Kelvin. At higher temperatures 
many partial waves contribute
and the multi-channel, close coupling theory described
in the previous sections is applied.
 We display results obtained for magnetic
fields that range between 1 and 4 Tesla. In the previous paragraphs
 we justified the neglect of the hyperfine interaction, in the calculations
for field strengths $ B > 1 T$.
The results shown in Fig. 5 are based on this approximation, but because
of the anisotropy in the dipolar interaction, this calculation  
still involves a large number of channels since coupled partial wave angular 
momenta up to $J \approx 20$ are required for convergence at temperatures
$ T > 1 K$. We find that the elastic cross sections
dominate and are largely insensitive to the value of the applied field.
In the range of applied fields considered and in the $\mu K$ collision energy region, 
the dipolar cross sections are about $10^{-3}$ smaller than the elastic
cross sections. The ratio decreases at higher collision energies and
at cryogenic temperatures the ratio of elastic to inelastic 
cross sections $ \approx 10^{-6}$. At higher temperatures we find that the 
dipolar relaxation cross sections decrease as the applied field is increased.
In Fig. 5, we note that the elastic cross sections tend to a constant value
as the gas temperature approaches the $\mu \, \rm K$ range whereas the relaxation cross
sections increase, in conformity with the Wigner threshold laws. Both the elastic
and inelastic cross sections display resonance features discussed
in the previous paragraphs.
 Because of the favorable ratio of
 elastic to inelastic cross sections, spin polarized sodium is, potentially, a good
 candidate for buffer gas loading and evaporative cooling.

\begin{acknowledgments}
This work was supported by an NSF grant to the MIT-Harvard Center
for Ultra cold Atoms (CUA). I thank the CUA for support as a
Visiting Scientist, and the
 Institute
for Theoretical Atomic and Molecular Physics (ITAMP) for their
hospitality while this work was undertaken. I thank Alex Dalgarno and Roman Krems
for a critical reading of this manuscript and for pointing out a numerical error 
in a previous version of this
manuscript. I also wish to thank John Doyle,
Jack Harris, Jonathan Weinstein and Wolfgang Ketterle for useful
discussions.
\end{acknowledgments}

\end{document}